# Optimal Design of Neural Network Structure for Power System Frequency Security Constraints

*Zhuoxuan Li, Zhongda Chu\*, Fei Teng*

*Department of Electrical & Electronic Engineering, Imperial College London, London, United Kingdom*
*\* z.chu18@imperial.ac.uk*

## Abstract

Recently, frequency security is challenged by high uncertainty and low inertia in power system with high penetration of Renewable Energy Sources (RES). In the context of Unit Commitment (UC) problems, frequency security constraints represented by neural networks have been developed and embedded into the optimisation problem to represent complicated frequency dynamics. However, there are two major disadvantages related to this technique: the risk of overconfident prediction and poor computational efficiency. To handle these disadvantages, novel methodologies are proposed to optimally design the neural network structure, including the use of asymmetric loss function during the training stage and scientifically selecting neural network size and topology. The effectiveness of the proposed methodologies are validated by case study which reveals the improvement of conservativeness and mitigation of computation performance issues.

## 1 Introduction

As achieving net zero is globally on the agenda, power systems are expected to increase the penetration of RES. However, characterised by high uncertainty and low inertia, RES are challenging to power system frequency security. In the context of UC problem, frequency security is traditionally considered by static approaches such as setting specific requirements for reserve, inertia or kinetic energy [1]. Due to the vague representations of frequency dynamics, static approaches usually leave excessive security margin, leading to significantly uneconomical dispatch plans.

With the help of mathematical models such as multi-machine swing equation, frequency security constraints can be constructed by analytical approaches with a focus on certain frequency security indices including frequency nadir and Rate of Change of Frequency (RoCoF) [2]. For example, [3] gives an expression of post-fault frequency trajectory, and its methodology is extended in [4] where damping effect is involved into consideration. In [5], a more detailed expression of post-fault frequency trajectory is derived regarding frequency dynamics as a closed-loop problem, and nonlinear expressions of frequency security indices are incorporated into UC through piecewise linearisation [6].

Although analytical approaches provide more realistic descriptions of frequency dynamics, the accuracy is limited by the following reasons. Firstly, it is almost impossible for analytical approaches to capture frequency at different buses and the frequency at Center Of Inertia (COI), discussed in multi-machine swing equation, may not be able to accurately represent frequency at different buses especially when the inertia distribution is uneven. Secondly, turbine dynamics and governor controllers contain highly nonlinear and nonconvex components, which is difficult to be incorporated into optimization problems analytically.

On the other hand, many data-driven techniques have been widely discussed for their applications in power system security analysis [7], and exploiting these techniques to formulate frequency security constraints is a rising research topic. This technique is initially introduced in [8] to identify frequency security conditions during microgrid islanding events. Later in [9], optimal classification trees, another data-driven approach, is utilised to predict frequency nadir, RoCoF, and quasi-steady state frequency, which is also integrated into UC problem with some manipulations. Similar philosophy to [7] is further extended to large-scale power systems and the high computational cost associated with the labelling bottleneck is mitigated by an active sampling algorithm [10]. Logistic regression has also been applied in constructing frequency security constraints [11] in the context of robust UC.

According to the comparison in [10], neural network has the best performance among many machine learning models in terms of prediction accuracy, and is thus the most promising model to fulfill the potential of data-driven approaches. However, there are few publications discussing how such neural networks should be designed. Actually, the application of neural network in frequency security formulation is challenged by at least two factors below. Primarily, the conservativeness is not rigorously analysed, leading to the risks of unacceptable constraint violations and security issues related. Besides, the computational cost of solving UC with neural network-embedded constraints is significantly high.

To deal with the challenges above, this paper studies the optimal design of neural network structure. Main contributions of this paper are twofold: 1) application of asymmetric loss function to obtain more conservative predictor, considering the cost sensitivity of frequency security prediction, and 2) determination of the neural network size to balance the trade-off between prediction accuracy and computation cost.

The rest of this paper is organised as follows. Section 2 introduces the formulation of the problem, i.e. UC constrained frequency security constrained based on neural network. Methodologies with respect to optimal structural design of neural networks are elaborated in Section 3. Section 4 shows the effectiveness of the proposed methods by a case study, followed by conclusions drawn in Section 5.

## 2  Problem Statement

Given the predicted outputs of RES, UC is usually formulated as a mixed integer linear programming (MILP) problem whose solution tells the optimal status and dispatch of synchronous generators. Consider a general day-ahead hourly dispatch scenario where the objective function, i.e., the total system operation cost is:

$$\min \sum_{t=1}^{T}\sum_{g=1}^{N_g}(\pi_g^F u_{g,t} + \pi_g^C P_{g,t}^G) + \sum_{t=2}^{T}\sum_{g=1}^{N_g}\pi_g^U v_{g,t}, \quad (1)$$

where $P_{g,t}^G$, $u_{g,t}$ and $v_{g,t}$ are respectively the active power output and binary variables denoting the online status and starting up state of generator $g$ at time step $t$; $\pi_g^C$, $\pi_g^F$ and $\pi_g^U$ are the associated cost. Conventional UC constraints related to power flow, power balance and voltage magnitudes, as well as minimum up and down time, power output and ramp of synchronous generators are omitted here and [8], [12], [13], can be referred for details.

Data-driven frequency security constraints are transformed from a frequency security predictor which predicts the value of certain frequency security index of a dispatch. For ease of demonstration and comparison, methodologies proposed in [10] is adopted here. Specifically, the predictor to be built is a neural network regressor that estimates frequency nadir (minimum of all buses) after the synchronous generator with maximum output power is suddenly disconnected from the network, namely 'N - 1' principle. Note that inputs of the predictor (features) must be derived from static decision variables of UC rather than transient quantities. Based on multi-machine swing equations, the following variables are selected to formulate the feature vector [10], [14]:

$$\boldsymbol{x}_t = [u_{1,t}^G,...,u_{N_g,t}^G,0,...,0,P_{g_{max},t}^G,0,...,0], \quad (2)$$

where $\boldsymbol{x}_t$ is the feature vector at time step $t$, $g_{max}$ is the index of synchronous generator with maximum active power output, i.e.,

$$g_{max,t} = \underset{g}{argmax}\, P_{g,t}^G \quad g=1,\ldots,N_g, \forall t. \quad (3)$$

It is apparent that there are $2N_g$ elements in the feature vector: the first $N_g$ elements are operating states of all synchronous generators, while the next $N_g$ elements are zero except the $N_g + g_{max}$-th element whose value is the maximum active power output of all synchronous generators, illustrated by (3). These elements are selected as they dominate the frequency trajectory after the disconnection of synchronous generator with maximum power output.

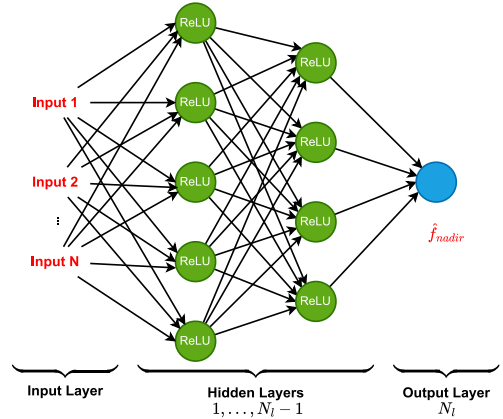

Fig. 1: Structure of neural network frequency security predictor

Multilayer perceptron, the most basic structure of neural network is applied for building frequency security predictor. Shown in Fig. 1, it is basically a linear model except for the nonlinear activation functions in the neurons of hidden layers. Fortunately, if ReLU, a special piecewise linear activation function defined as (4), is adopted, the entire network can be linearised and transformed into linear constraints with the manipulations in [15]. In fact, linear expressions (5a)-(5e) are equivalent to (4) where $\underline{h}$ and $\overline{h}$ satisfy $\underline{h} < Z < \overline{h}$ and $\underline{h} < 0 < \overline{h}$, with $\underline{h}$ and $\overline{h}$ being the big-M constants.

$$z = \text{ReLU}(Z) = \max\{0, Z\} = \begin{cases} 0 & Z<0 \\ Z & Z \geq 0 \end{cases} \quad (4)$$

$$z \leq Z - \underline{h}(1-a) \quad (5a)$$
$$z \geq Z \quad (5b)$$
$$z \leq \overline{h}a \quad (5c)$$
$$z \geq 0 \quad (5d)$$
$$a \in \{0,1\} \quad (5e)$$

In this way, activation function ReLU can be transformed into linear constraints for a MILP problem. The definition of $g_{max,t}$ in (3) is also linearised as follows [10]:

$$P_{g_{max},t}^G - P_{g,t}^G \leq \Gamma(1-\mu_{g,t}) \quad g=1,\ldots,N_g, \forall t \quad (6a)$$

$$\sum_{g=1}^{N_g} \mu_{g,t} = 1 \quad \forall t, \quad (6b)$$

where $\mu_{g,t}$ is an auxiliary binary variable and $\Gamma$ is also a big-M constant which is always greater than the maximum output power of all synchronous generators. Other components of the neural network predictor are already linear, and the complete frequency security constraints are listed below.

$$\mu_{g,t}, \mu_{g_{max},t} \in \{0,1\} \quad g, g_{max} = 1,\ldots,N_g, \forall t \quad (7\,a)$$

$$\mathbf{x}_t(g) = u_{g,t} \quad g = 1,\ldots,N_g, \forall t \quad (7\,b)$$

$$P^G_{g_{max},t} - P^G_{g,t} \leq \Gamma(1-\mu_{g,t}) \quad g, g_{max} = 1,\ldots,N_g, \forall t \quad (7\,c)$$

$$\sum_{g=1}^{N_g} \mu_{g,t} = 1 \quad \forall t \quad (7\,d)$$

$$\mathbf{x}_t(N_g + g) - P^G_{g,t} \geq -\Gamma(1-\mu_{g,t}) \quad g = 1,\ldots,N_g, \forall t \quad (7\,e)$$

$$\mathbf{x}_t(N_g + g) - P^G_{g,t} \leq \Gamma(1-\mu_{g,t}) \quad g = 1,\ldots,N_g, \forall t \quad (7\,f)$$

$$0 \leq \mathbf{x}_t(N_g + g) \leq \Gamma\mu_{g,t} \quad g = 1,\ldots,N_g, \forall t \quad (7\,g)$$

Feature vectors are constrained by (7a) - (7g) where $\mathbf{x}_t(m)$ is the $m$-th element of feature vector $\mathbf{x}_t$.

$$Z_{1,n,t} = \mathbf{w}_{1,n,t} \cdot \mathbf{x}_t + b_{1,n,t} \quad \forall n,t \quad (8\,a)$$

$$Z_{l,n,t} = \mathbf{w}_{l,n,t} \cdot \mathbf{z}_{l-1,t} + b_{l,n,t} \quad l = 2,\ldots,N_l, \forall n,t \quad (8\,b)$$

$$z_{l,n,t} \leq Z_{l,n,t} - \underline{h}_{l,n}(1-a_{l,n,t}) \quad l = 1,\ldots,N_l-1, \forall n,t \quad (8\,c)$$

$$z_{l,n,t} \geq Z_{l,n,t} \quad l = 1,\ldots,N_l-1, \forall n,t \quad (8\,d)$$

$$z_{l,n,t} \leq \overline{h}_{l,n} a_{l,n,t} \quad l = 1,\ldots,N_l-1, \forall n,t \quad (8\,e)$$

$$z_{l,n,t} \geq 0 \quad l = 1,\ldots,N_l-1, \forall n,t \quad (8\,f)$$

$$a_{l,n,t} \in \{0,1\} \quad l = 1,\ldots,N_l-1, \forall n,t \quad (8\,g)$$

$$\underline{y} \leq Z_{N_l,1,t} \quad \forall t \quad (8\,h)$$

(8a)-(8g) express the linearised neural networks. For each time step $t$, a neural network with $N_l$ layers is built to predict frequency nadir after the disconnection of the generator with maximum power output. For every time step $t$, $Z_{l,n,t}$, $z_{l,n,t}$, $\mathbf{w}_{1,n,t}$ and $b_{l,n,t}$ are respectively the input, output, weight vector and bias of the $n$-th neuron at $l$-th layer. $\underline{y}$ is the minimum permissible value of frequency nadir.

## 3 Methodologies

*3.1 Conservativeness Improvement by Using Asymmetric Loss Functions*

The training process of a neural network predictor minimises the gap between predicted and actual values, which is done, in common training algorithms, by minimising the value of a loss function. Absolute value loss and quadratic loss are the most widely used loss functions, respectively shown in (9) and (10) where $y$ is actual value and $\hat{y}$ is the corresponding predicted value [16].

$$L(y,\hat{y}) = |y-\hat{y}| \quad (9)$$

$$L(y,\hat{y}) = (y-\hat{y})^2 \quad (10)$$

According to regularisation types [17], absolute value loss is categorised as L1 loss functions while quadratic loss is categorised as L2 loss functions. Meanwhile, both absolute value loss and quadratic loss are symmetric functions, indicting that errors are treated non-directionally such that overestimation and underestimation weight the same during the minimization of loss.

It is appropriate to use symmetric loss functions discussed above in most applications of regression. However, using symmetric loss functions could be problematic in cost sensitive situations, where positive and negative errors do not have the same effect in practice [18]. For example, investors in financial market are characterised by risk aversion, i.e., the utility of losing certain amount of money is not the exact opposite of the utility of winning the same. The existence of difference between positive and negative error is exactly the meaning of cost sensitiveness issue. Prediction of frequency security is a similar scenario of cost sensitiveness. Overconfidence about system strength could lead to under frequency load shedding (UFLS), which is much more expensive than the extra cost paid by system operators if frequency security condition is underestimated. As a result, frequency security predictor are expected to be conservative respecting to the cost sensitive characteristics.

Asymmetric loss functions are classic solution to cost sensitiveness issues. Similar to L1 absolute value loss and quadratic loss, variants of asymmetric loss functions are shown in (11) and (12), belonging respectively to L1 and L2 loss functions [18], displayed in Fig. 2.

$$L(y,\hat{y}) = \begin{cases} C^+ |\hat{y}-y| & \hat{y} \geq y \\ C^- |\hat{y}-y| & \hat{y} < y \end{cases} \quad (11)$$

$$L(y,\hat{y}) = \begin{cases} C^+ (\hat{y}-y)^2 & \hat{y} \geq y \\ C^- (\hat{y}-y)^2 & \hat{y} < y \end{cases} \quad (12)$$

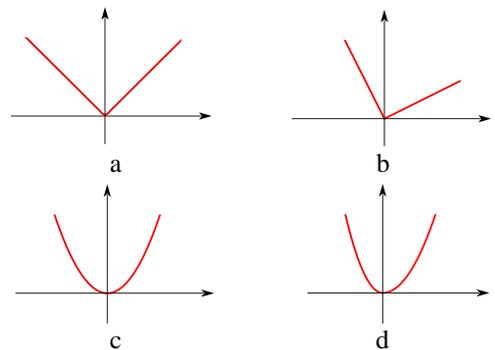

Fig. 2: Symmetric and asymmetric loss function
(a) MAE loss  (b) lin-lin loss
(c) MSE loss  (d) quad-quad loss

It is obvious that factors $C^+$ and $C^-$ are involved to make a predictor conservative or aggressive. One can let $C^+ > C^- > 0$ so that overestimation of frequency nadir, namely $\hat{y} > y$ will be punished more during the training process, and the conservativeness can be controlled by adjusting factors $C^+$ and $C^-$. Note that the improved conservativeness will deteriorate the prediction accuracy, which is discussed later in Section 4.

*3.2 Design of Neural Network Size and Topology*

Size and topology of the neural network not only determine the accuracy of the predictor, but also determine the computation cost of solving the UC with neural network constraints. According to formulations in Section 2, each neuron is designated an auxiliary binary variable presenting its activation state. Hence, the numbers of integer variables and constraints increase dramatically as the size of the neural network grows. This could lead to poor tractability in computation especially when solving large-scale problems like UC, which is admitted in [19], [20].

It is necessary to take both accuracy and computational tractability into consideration when designing neural network size and topology. Unfortunately, such designing mainly relies on experiments and empirical observations. The effects of neural network sizes and topology are studied in two separate experiments: one with fixed topology but varying sizes and another with fixed size but varying topology. Accuracy and computation performance are recorded and compared in Subsection 4.3

# 4 Case Study

*4.1 Dataset Generation*

A dataset of post fault frequency trajectories is essential to construct data-driven frequency security constraints. New England 39-bus system, a benchmark for power system stability studies [21], is modified to demonstrate a power system with high RES penetration. Specifically, four 900MW wind farms consisting of doubly-fed induction generators (DFIGs) are penetrated respectively at bus 2, 10, 20 and 25, which is displayed in Fig. 3. Parameters of this system can be found in [22].

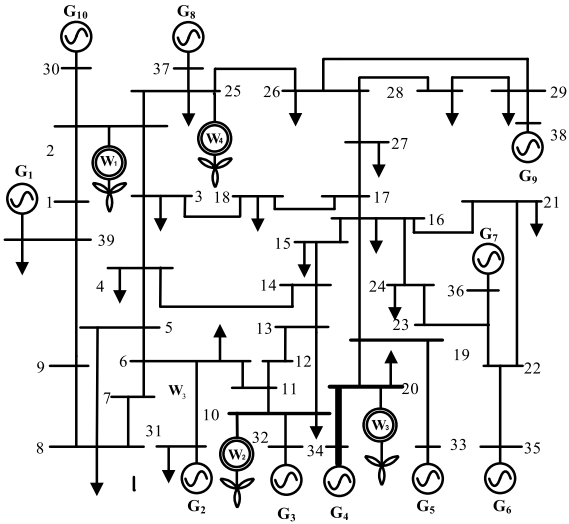

Fig. 3: Diagram of the Modified New England 39-bus Power System

Dynamic simulation for generating dataset is performed on MATLAB/Simulink R2021a. Full-order model of synchronous generators are implemented in simulations while wind farms are modelled as aggregated DFIGs. Because the contingency is always the disconnection of the synchronous generator with maximum power output, post fault frequency trajectories are determined by the steady state operating conditions before the contingency.

Generated samples are setup to simulate post fault frequency trajectories after the loss of synchronous generator with maximum active power output. After filtering out unconverged trajectories, every sample are labelled with its minimum measured frequency nadir collected at all buses. Fig. 4 summarise the procedure of dataset generation.

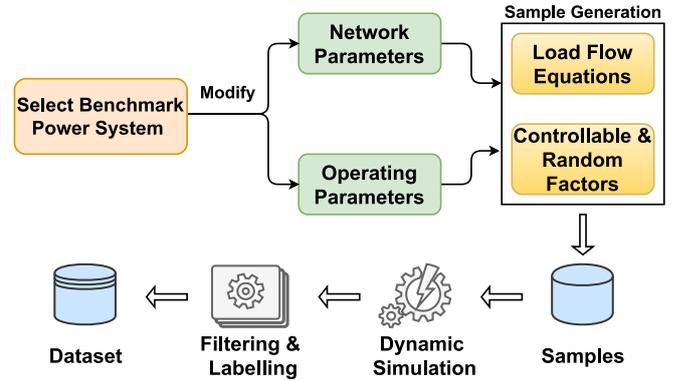

Fig. 4: Procedure of Dataset Generation

*4.2 Solution of UC*

Totally 1676 samples are generated and simulated, and the obtained dataset is used to train the neural network predictor. To validate the performance of different loss functions, neural network frequency security predictors with 256 neurons in single hidden layer, but different loss functions are trained by Tensorflow 2.4.0. Table 1 compares the performances of these predictors, where MAE means absolute value loss. It can be observed that L2 loss functions slightly outperforms L1 ones in terms of prediction accuracy. Moreover, the increase of conservativeness leads to decreased accuracy so that a compromise should be made between conservativeness and accuracy.

Table 1: Performance of neural network frequency security predictors trained with different loss functions

| Loss Function Type | Asymmetric Cost Ratio $C^+/C^-$ | Proportion of Conservative Prediction | MAE (Hz) | $R^2$ |
|---|---|---|---|---|
| L1 | 1(symmetric) | 48.98% | 0.0337 | 0.9588 |
| L1 | 5 | 65.31% | 0.0351 | 0.9553 |
| L2 | 1(symmetric) | 48.35% | 0.0334 | 0.9651 |
| L2 | 5 | 69.16% | 0.0436 | 0.9599 |

Frequency security constrained UC is ready to be solved after the neural network predictor is transformed into constraints.

In a UC problem modelling 24-hour ahead dispatch scheduling, total active load is setup to vary between 3GW to 7GW. Frequency security criterion is whether frequency drops further than 0.8Hz from 50Hz normal frequency. Fig. 5 presents the solutions of UC in the following three cases:

    a. no frequency security constraints
    b. symmetric loss $C^+/C^- = 1$
    c. asymmetric loss $C^+/C^- = 5$

In Fig. 5, x-axis denotes the index of synchronous generator and the y-axis denotes the hourly time step. Uncommitted synchronous generators are presented as circles with empty interior while committed ones are marked as circles with fulfilled interior. Moreover, the opacity of a filled circle indicates the ratio between committed active power and maximum active power of this generator.

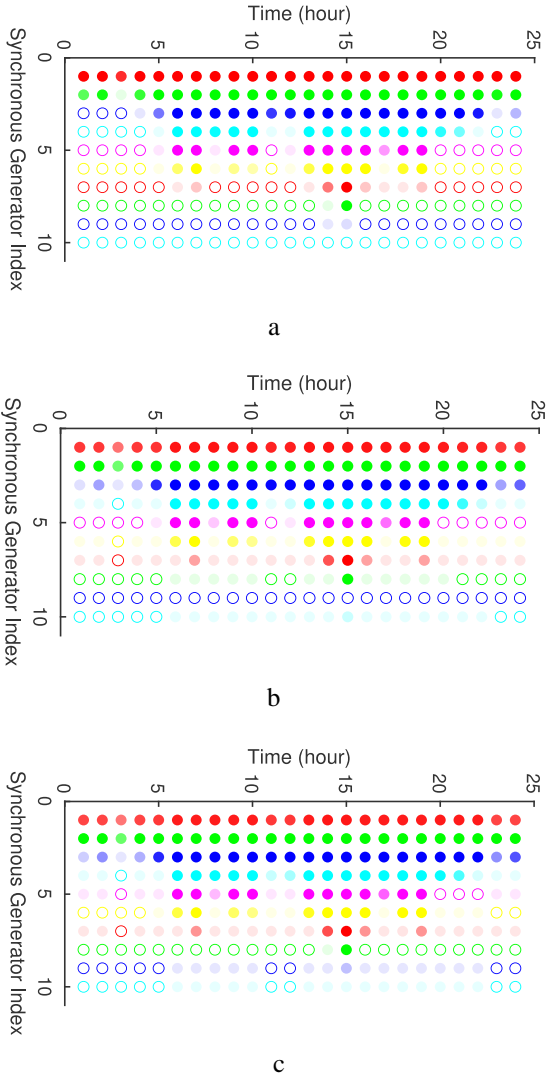

Fig. 5: Results of UC in three different cases
(a) no frequency security constraints
(b) symmetric loss $C^+/C^- = 1$
(c) asymmetric loss $C^+/C^- = 5$

In Fig. 5, x-axis denotes the index of synchronous generator and the y-axis denotes the hourly time step. Uncommitted synchronous generators are presented as circles with empty interior while committed ones are marked as circles with fulfilled interior. Moreover, the opacity of a filled circle indicates the ratio between committed active power and maximum active power of this generator.

Compared with case (a), there are much more committed synchronous generators in case (b), significantly increasing the frequency security margin. Similar dispatch planning is witnessed in the result of case (c), yielding larger frequency security margin. Consider the setups of the cases listed above, data-driven frequency security predictor based on neural network has successfully guided the dispatch planning, and the proposed application of asymmetric loss function does bring more conservative results.

### 4.3 Computation Performance

The UC problem is solved on a PC with AMD Ryzen 5 2500U 2.00 GHz CPU and 8GB RAM. Dual simplex algorithm is deployed on Gurobi, a common solver for MILP. The branch and cut algorithm consecutively searches for new lower upper bounds and higher lower bounds throughout the solving process. Every upper bound is linked with a potential optimal solution, and a lower bound is a validated minimum of the objective function. The solving process finishes when the gap between the lowest upper bound and the highest lower bound, presented by MIPGap, shrinks to zero. Let all neurons except the one in output layer placed in a single hidden layer, Table 2 shows the relationship between accuracy and computation performance for different neural network sizes.

Large neural network size improves the accuracy, but the computation performance becomes undesirable when there 8 or more neurons. Actually, when there are more 32 neurons in hidden layers, it takes hours to completely solve the UC problem. The required time for completely solving UC surges with the neural network size, such that only a suboptimal solution, or a not completely validated optimal solution, can be found within 1000s. Therefore, rather than using large-size neural networks, small-size neural networks are preferred especially when the accuracy is good enough. For example, an acceptable criterion can be MAE < 0.05Hz which needs only 32 neurons.

Table 2: Relationship between accuracy and computation performance for different neural network sizes

| Neurons | MAE | $R^2$ | Computation Time | MIP Gap |
|---|---|---|---|---|
| 2 | 0.2213 | 0.0026 | 5s | 0.00% |
| 4 | 0.0512 | 0.9150 | 46s | 0.00% |
| 8 | 0.0495 | 0.9223 | 1000s | 0.48% |
| 32 | 0.0404 | 0.9487 | 1000s | 3.06% |
| 128 | 0.0308 | 0.9649 | 1000s | 5.47% |

Likewise, Table 3 presents how topology influences prediction accuracy and computation time, based on a neural network with 32 neurons placed in one or more hidden layers.

Table 3: Accuracy and computation performance for different neural network topology.

| Topology | MAE | $R^2$ | Computation Time | MIP Gap |
|---|---|---|---|---|
| [32] | 0.0404 | 0.9487 | 1000s | 3.06% |
| [16,16] | 0.0392 | 0.9538 | 1000s | 7.21% |
| [24,8] | 0.0420 | 0.9473 | 1000s | 6.15% |
| [8,24] | 0.0379 | 0.9612 | 1000s | 4.73% |
| [16,8,8] | 0.0404 | 0.9520 | 1000s | 3.64% |
| [8,16,8] | 0.0369 | 0.9573 | 1000s | 6.58% |
| [16,12,4] | 0.0385 | 0.9561 | 1000s | 7.44% |
| [12,16,4] | 0.0379 | 0.9598 | 1000s | 7.07% |

It is obvious that single hidden layer topology has smaller MIPGap after 1000s. In fact, decision variables in multi-layer topology are much more coupled than that in single layer topology. Illustratively, stronger coupling requires more computation efforts, and hence single hidden layer topology is the most computationally efficient.

The experimental result comply with the expressions of frequency security constraints constructed by neural networks. It can be observed from (8a)-(8h) that the number of constraints is determined and only determined by the number of neurons. Hence, the size of neural network plays a key role in computation performance. Besides, the topology of neural networks determine the number of parameters in a neural network and to what extent these variables are coupled, and can also influence the computation performance.

Although the single hidden layer topology may not be the most accurate, other topology does not significantly outperform it. Similar phenomena are also observed in neural network with other sizes, so single hidden layer is recommended for this particular issue in practice.

## 5 Conclusions

This paper discusses the structural design of neural network as frequency security constraints. The prediction of frequency security is considered as a cost sensitive problem, and the high computational cost of solving UC with delicate neural network constraints is identified. Accordingly, methodologies are proposed to tackle these challenges when designing the neural network. Asymmetric loss function is proposed to be applied during the training of neural network to improve the conservativeness of the predictor and transformed frequency security constraints. Meanwhile, proper sizing the neural network can mitigate the poor computational performance by reducing the number of integer variables and constraints. The effectiveness of the proposed methodologies are validated by the case study, which reveals that asymmetric loss function can indeed yield more conservative dispatch. When maximum accuracy is achieved, the computation performance of solving UC is poor. However, reducing the size of neural network and adopting single hidden layer topology are proved to be practical methods to decrease computational cost while maintaining the accuracy at a satisfactory level.